\documentstyle[11pt]{article}

\title{Analytic approach to the complete set of QED corrections
        to fermion pair production \\in \ee - annihilation}
 
\author{D. Bardin, M. Bilenky, A. Chizhov, A. Sazonov\\
          Joint Institute for Nuclear Research, Dubna,
  \\ Head Post Office P.O. Box 79, SU - 101000 Moscow, USSR
                  \\   \\ O. Fedorenko
\\ Petrosavodsk State University, SU-185018 Petrosavodsk, USSR
\\ \\and \and T. Riemann, M. Sachwitz
    \\ Institut f\"ur Hochenergiephysik,
    \\   Platanenallee 6, DDR-1615 Zeuthen/Brandenburg\thanks{Address to be
expected after 3 October 1990: O-1615 Zeuthen/Brandenburg(FRG)}}
\date{18 March 1990}
 
\newcommand{\ee}{e$^{+}$e$^{-}$}
\newcommand{\afb}{$A_{FB}\:$}
\newcommand{\st}{$\sigma_{T}\:$}
\newcommand{\oalf}{O($\alpha$)$\:$}

\begin{document}
 
\maketitle
\vspace{1.cm}
   ABSTRACT
\vspace{1.cm}
\nopagebreak
 
     We   present   the  convolution  integral  for
       fermion pair production  in  the  electroweak  standard  theory
 to order \oalf  including   also  soft  photon  exponentiation.
       The  result  is complete  in the sense  that  it  includes
     initial  and  final  state      radiation   and
        their  interference.  From the basic result - analytic
 formulae for the  differential      cross section - we also derive
the corresponding  expressions for total cross section \st
and integrated forward-backward asymmetry  \afb.
The numerical importance of different contributions for the
 analysis of  experiments  at  LEP/SLC energies is discussed.
 
\vspace{.8cm}

\vfill\eject
1.      Introduction (3)
 
2.      Notations and phase space parametrisation (3)
 
3.      Initial state radiation (7)
 
3.1.    The radiator functions for the angular distribution (7)
 
3.2.    Soft photon exponentiation (9)
 
3.3.    Integrated cross section and asymmetry (9)
 
4.      Initial-final state interference radiation (13)
 
4.1.    The radiator functions for the angular distribution (13)
 
4.2.    Integrated cross section and asymmetry (16)
 
5.      Final state radiation (19)
 
5.1.    The radiator functions for the angular distribution (19)
 
5.2.    Integrated cross section and asymmetry to order \oalf (20)
 
5.3     Soft photon exponentiation (20)
 
6.      Discussion (22)
 
\vfill\eject
   INTRODUCTION
\vspace{1.cm}
\nopagebreak
 
     Stimulated by the possibility of very precise experiments
     at \ee - storage rings, the QED corrections for the reaction
 
\begin{equation}
e^{+}(k_{2}) + e^{-}(k_{1})
 \longrightarrow ( \gamma ,Z) \longrightarrow f^{+}(p_{2})
+ f^{-}(p_{1}) + n\gamma(p)
\vspace{.5cm}
\label{eq:cross}
\end{equation}
have  been  studied in very detail for a large  energy  region
including  the  Z resonance.  A comprehensive  collection  and
comparison   of  available  results  within  the   electroweak
standard  theory $^{\cite{no:gws}}$
 may be found in $^{\cite{yr:alt}}$
 and  in  references
quoted therein.  Analytic and semi-analytic formulae proved to
be  of great value both for a deeper understanding of  process
(~\ref{eq:cross}) and for ensuring a high reliability of
Monte Carlo codes  $^{\cite{yr:kle}}$.
 Recently, the analytic approach to QED has also been used
for  the analysis of Z line shape data at  LEP/SLC
 $^{\cite{pl:lep}}$. The
line  shape is the integrated total cross section \st(s) as
function  of  $s = 4E^{2}$ where $E$ is the beam  energy.  This  most
inclusive observable has been studied theoretically since many
years  and  is well-described to order \oalf
 $^{\cite{np:bom}-\cite{np:gps}}$ with  soft
photon  exponentiation  $^{\cite{np:gps,pr:cbc}}$
 and also for  several  massive
intermediate  bosons  $^{\cite{pl:lrs}}$
,  higher order leading  logarithmic
corrections  $^{\cite{np:kfa,pl:nit}}$
 and the complete QED initial state
radiation  corrections to order O($\alpha^{2}$)$\:^{\cite{np:bbn}}$.
A compilation of  many
 important contributions is contained in $^{\cite{yr:fab}}$ .
 
     Only   recently,   similar  analytic  results  have  been
     published  for the integrated forward-backward asymmetry \afb .
The  complete \oalf convolution integral with soft  photon
exponentiation  for \afb  was obtained in  $^{\cite{pl:dyb}}$
(see also  $^{\cite{du:bbf}}$ )  and
the leading logarithmic approximation for initial state radiation
including O($\alpha^{2}$)  terms in  $^{\cite{pr:bbn}}$
. A survey on results about \afb is contained in  $^{\cite{yr:boh}}$ .
 
     Compared  to \st  and  \afb, analytic  results  for  the
differential cross section d$\sigma$/dcos$\Theta$ are scarce. Earlier
attempts  are  $^{\cite{np:kum,np:pas}}$
.  Other distributions are treated in  $^{\cite{np:ins}}$ .
In   pure QED,  the first compact analytic expressions for  hard
bremsstrahlung  corrections to the differential cross  section
have   been derived in  $^{\cite{ap:fer}}$
. A formalism for leading logarithmic
approximations may be found in  $^{\cite{pr:bbn}}$ , though without
application to d$\sigma$/dcos$\Theta$. Recently, semi-analytic
formulae allowing
for quite realistic cuts have been applied in  $^{\cite{pl:bss}}$ .
     This  article contains the systematic presentation of  an
     analytic   calculation  of  QED  corrections  to  the  angular
distribution for reaction  (~\ref{eq:cross}) .
Though computer codes relying
on  the  present results have already been released  and  used
 $^{\cite{l3:ade}}$
  for  applications  in LEP  physics,  very  few  of  the
material   has  been  presented  so  far
 $^{\cite{pr:bfr}}$
;   see  also  $^{\cite{yr:kle,yr:fab,yr:boh}}$
. In chapter 2 we introduce the notation and describe
in  short two derivations of the analytic expressions for  the
angular distribution.  Chapter 3 to 5 contain the basic result
which   consists of compact,  explicit expressions for the  soft
and hard photon radiator functions in the convolution integral
due    to initial (chapter 3) and final (chapter 5) state
radiation and their interference (chapter 4) to order \oalf.
 They also   include higher order corrections and some formulae
for \st and \afb. Chapter 6 contains numerical results and conclusions.
 
 
\vspace{1.cm}
     2. NOTATIONS AND PHASE SPACE PARAMETRISATION
\vspace{1.cm}
\nopagebreak
 
     The explicit analytic formulae for the differential
     cross section proved to be too cumbersome to be
described in detail here.  For that reason,  we choose a semi-
analytic  presentation with the following notation:
\begin{equation}
\frac{d\sigma}{dc} = \sum_{m,n=0,1} \;\;\sum_{A=T,FB} \;\;\sum_{a=e,i,f}
                    Re[\sigma_{A}^{a,0}(s,s;m,n)\:R_{A}^{a}(c;m,n)].
\label{eq:d21}
\end{equation}
 
The scattering angle $c = \cos\Theta$ is defined between the produced
fermion $f^{+}(f = \mu,\nu,q)$ and  the  positron  beam.  The functions
$\sigma_{A}^{a,0}(s,s;m,n)$ are reduced Born cross sections defined below
and  the  corrections $ R_{A}^{a}(c;m,n)$ can be  expressed  by
semi-analytic  formulae for soft and  hard parts of QED bremsstrahlung
radiator functions $S_{A}^{a}(c,\epsilon;m,n)$ and $H_{A}^{a}(v,c)$:

\begin{equation}
R_{A}^{a}(c;m,n) = \int_{0}^{\Delta} dv
\frac{\sigma_{A}^{a,0}(s,s';m,n)}  {\sigma_{A}^{a,0}(s,s;m,n)}
 [\delta(v) S_{A}^{a}(c,\epsilon;m,n)
                   + \theta(v-\epsilon) H_{A}^{a}(v,c)].
\label{eq:r22}
\end{equation}
 
In (~\ref{eq:r22}), we allow for a possible cut-off $\Delta$ (
  0$ < \Delta \leq$ 1-4m$_{f}^{2}$/s)
on  the  energy $v$ of the emitted photon in units of  the  beam
energy. The effective energy $s'$ of the created fermion pair is
$s' = (1-v) s$. The sums over $m,n$ in (~\ref{eq:d21}) include photon $(m = 0)$
and Z boson $(m = 1)$ exchange.  A generalisation  to the case  of
additional vector bosons Z$_{n},  n>$1, is straightforward. Indices
$a = e,f,i$  are  used for initial and final state  radiation  and
their interference.  The angular dependence  of the Born cross
section has been formally included into the initial state contribution
$a = e$. The P-even and
P-odd  cross  section parts carry index $A = T$ and $A = FB$.
Under CP-invariance, they are also C-even (C-odd), correspondingly
. After symmetric  (anti-symmetric)
integration  over  cos$\Theta$
 the $R_{T}^{a}$- functions ($R_{FB}^{a}$- functions) yield the
 total cross section \st
(the integrated forward-backward asymmetry \afb ):
\begin{equation}
   \sigma_{T} = \int_{-1}^{1} dc \frac{d\sigma}{dc},
\label{eq:s23}
\end{equation}
\begin{equation}
A_{FB} = \frac{\sigma_{FB}}{\sigma_{T}}= \frac{1}{\sigma_{T}}
    \left[\int_{0}^{1} dc \frac{d\sigma}{dc} - \int_{-1}^{0}dc
    \frac{d\sigma}{dc}\right].
\label{eq:a24}
\end{equation}
Within our terminology,
\begin{equation}
   \sigma_{A} = \sum_{m,n=0,1} \;\sum_{a=e,i,f} Re [d_{A} \:
    \sigma_{A}^{a,0}(s,s;m,n) \:R_{A}^{a}(m,n)],
\label{eq:s25}
\end{equation}
\begin{equation}
R_{A}^{a}(m,n) = d_{A}^{-1} \int_{0}^{1} dc \:R_{A}^{a}(c;m,n),
\hspace{1cm}A = T,FB,
\label{eq:r26}
\end{equation}
\begin{equation}
d_{T} = \frac{4}{3}, \hspace{1cm} d_{FB} = 1.
\end{equation}
 
The  additional numerical factors in (~\ref{eq:s25},~\ref{eq:r26}) ensure
the  usual  total
cross section normalisations for \st and $\sigma_{FB}$.
 
    The  dependence  of d$\sigma$/d$c$ on weak  neutral  couplings  and
electric charges of the fermions and on possible polarisations
and  weak loop corrections together with the typical Born-like
resonance  behaviour have been collected in the explicit  Born
factors $\sigma_{A}^{a,0}(s,s;m,n)$.   As   a  consequence,
  the   functions $R_{A}^{a}(c;m,n)$ depend only on the
 Z boson mass $M_{Z} = M_{1}$  and width  $\Gamma_{Z} = \Gamma_{1}$
 $^{\cite{np:abr,yr:fab}}$
  and on kinematic variables $s$, cos$\Theta$, and the energy
cut-off $\Delta$.  Of course,  fermion masses which are assumed
here to  be  small  compared  to $s, M_{Z},\Gamma_{Z}$ show
up  in  certain logarithmic mass singularities.
 
The  reduced  Born  cross  sections $\sigma_{A}^{a,0}(s,s';m,n)$
have different energy dependence for $a = e,i,f$:
\begin{equation}
   \sigma_{A}^{e,0}(s,s';m,n) = \sigma_{A}^{0}(s',s';m,n),
\end{equation}
\begin{equation}
   \sigma_{A}^{i,0}(s,s';m,n) = \sigma_{\bar{A}}^{0}(s,s';m,n),
\label{eq:s09}
\end{equation}
\begin{equation}
   \sigma_{A}^{f,0}(s,s';m,n) = \sigma_{A}^{0}(s,s;m,n).
\end{equation}
Here, $\bar{A} = FB$ or $T$  if  $A = T$ or $FB$, and
\begin{equation}
   \sigma_{A}^{0}(s,s';m,n) = \frac{\pi\alpha^{2}}{2 \:s}C_{A}
   (m,n;\lambda_{1},\lambda_{2},h_{1},h_{2}) c_{f} \frac{1}{2}
   [\chi_{m}(s')\chi_{n}^{*}(s) + \chi_{m}(s)\chi_{n}^{*}(s')],
\label{eq:s10}
\end{equation}
\begin{equation}
   \chi_{n}(s) = \frac{g_{n}^{2}}{4\pi\alpha}\: \frac{s}{s-m_{n}^{2}},
\label{eq:c211}
\end{equation}
\begin{equation}
   m_{n}^{2} = M_{n}^{2} - i M_{n} \Gamma_{n}(s),
\end{equation}
\begin{equation}
   \Gamma_{n}(s) \simeq \frac{s}{M_{n}^{2}} \Gamma_{n}.
\label{eq:g213}
\end{equation}
In (~\ref{eq:s10}),  the fermions have a color factor $c_{f}$
  = 3  in   case
of  quarks  and  $c_{f}$ = 1 for leptons.  Their vector  and  axial
vector  couplings  $v_{f}(n)$ and $a_{f}(n)$ to gauge boson Z$_{n}$ are
contained in $C_{A}$:
\begin{eqnarray}
   \lefteqn{
 C_{T}(m,n;\lambda_{1},\lambda_{2},h_{1},h_{2}) = } \nonumber \\
&  \{\lambda_{1}[v_{e}(m)v_{e}^{*}(n)   + a_{e}(m)a_{e}^{*}(n)]   +
&    \lambda_{2}[v_{e}(m)a_{e}^{*}(n)   + v_{e}^{*}(n)a_{e}(m)]\}
    \nonumber \\
&  \{h_{1}[v_{f}(m)v_{f}^{*}(n)   + a_{f}(m)a_{f}^{*}(n)]   +
&    h_{2}[v_{f}(m)a_{f}^{*}(n)   + v_{f}^{*}(n)a_{f}(m)]\},
\label{eq:c14}
\end{eqnarray}
\begin{equation}
  C_{FB}(m,n;\lambda_{1},\lambda_{2},h_{1},h_{2}) =
  C_{T}(m,n;\lambda_{2},\lambda_{1},h_{2},h_{1}).
\label{eq:c17}
\end{equation}
We  allow for longitudinal polarisations $\lambda_{-},\lambda_{+}$
  of both the
electron  and   positron  beams  and  for  final  states  with
helicities  $h_{-}, h_{+}$.  Due to the CP-invariance of the problem
and  the $(v,a)$ structure of the  interactions,  the  following
combinations of polarisations are possible:
\begin{equation}
  \lambda_{1} = 1 - \lambda_{+}\lambda_{-}, \hspace{1cm}
  \lambda_{2} =  \lambda_{+} - \lambda_{-},
\end{equation}
\begin{equation}
  h_{1} = \frac{1}{4}(1 - h_{+}h_{-}), \hspace{1cm}
  h_{2} = \frac{1}{4}(h_{+} - h_{-}).
\label{eq:h217}
\end{equation}
In  case of the standard theory  $^{\cite{no:gws}}$
,  we use the  conventional couplings of photon and Z boson:
\begin{equation}
  g_{0} = e, \hspace{1cm}v_{f}(0) = Q_{f},\hspace{1cm}a_{f}(0) = 0,
\label{eq:tr1}
\end{equation}
\begin{equation}
  g_{1} = (\sqrt{2} G_{\mu} M_{Z}^{2})^{1/2}, \hspace{1cm}
  v_{f}(1) = I_{3}^{L}(f) - 2 Q_{f} \sin^{2}\Theta_{W},
  \hspace{1cm}a_{f}(1) = I_{3}^{L}(f),
\label{eq:tr2}
\end{equation}
where  $I_{3}^{L}(f)$  is the third component of the weak  isospin  of
fermion $f$; 2 $I_{3}^{L}(e) = Q_{e}$ = -1.
 
     The inclusion of non-QED weak loop effects (see
 $^{\cite{pl:bbh,cp:bar,fp:hol,yr:alt}}$
and references quoted therein) is trivial if one uses the form
factor approach for their description
 $^{\cite{cp:bar}}$ . Coupling constants
$g_{0}, g_{1}$,  and the vector couplings $v_{e}(1), v_{f}(1)$
  become complex and
$s$-dependent.  Further details are explained in
 $^{\cite{fp:hol,yr:boh}}$. For an
exact treatment, an additional form factor $v_{ef}(1)$ replaces the
product  $v_{e}(1) v_{f}(1)$ in (~\ref{eq:c14})
 whenever it appears  there.  In
principle,  energy-dependent form factors should be understood
to  be  part of the integrand in (~\ref{eq:d21},~\ref{eq:r22})
.  We have checked  by
explicit  comparisons that,  due to the minor
s-dependence  of  them,  the numerical error which is implied by
the approximation is practically negligible
 $^{\cite{ph:bar,yr:kle}}$ .
 
For notational convenience,  all angular dependences have
been  included  into  the  radiator functions.  The  functions
$R_{A}^{a}(c;m,n)$
of (~\ref{eq:r22}) are result of an incoherent sum of real and virtual
photonic corrections. They are obtained by straightforward but lengthy
Feynman  diagram calculations with extensive use of the analytic
manipulation  programs SCHOONSCHIP   $^{\cite{cp:str}}$
and REDUCE    $^{\cite{rp:hea}}$. The $R_{A}^{a}(c;m,n)$ are
gauge-invariant and,  if integrated without cuts,  also
Lorentz-invariant.  Determinations of the vertex and box  corrections
  $^{\cite{np:bcf}}$,  soft  photon  and hard  photon  contributions
deserve different techniques, correspondingly.
 
     We  would like to give short comments on the two  methods
which  have been used for the \oalf bremsstrahlung phase  space
integrations.  In one approach
, we used the so called $R_{\gamma}$ -
system,  the  rest system of ($f^{-}\gamma$),  where the  three-momentum
relation  $\vec{p} (f^{-}$)  + $\vec{p} (\gamma$) = 0 is  fulfilled.
  The  phase  space parametrisation chosen is:
\begin{equation}
  \int d\Gamma = \frac{\pi^{2}}{4s^{3}}\int_{-c_{m}}^{c_{m}}
  d \cos \Theta
  \int_{0}^{s} x dx
  \frac{s-x}{s-x+m_{f}^{2}}\:\frac{1}{4\pi} \int_{-1}^{1}
  d \cos\Theta_{\gamma}^{R} \int_{0}^{2\pi} d \varphi_{\gamma}^{R},
\label{eq:idg}
\end{equation}
where ($\Theta_{\gamma}^{R},\varphi_{\gamma}^{R}$)
are photon angles in the $R_{\gamma}$ - system and $x$ the
momentum  fraction  of  $f^{+}$ in the cms in units  of  the  beam
energy:
\begin{equation}
   x = -2 p (f^{+}) [ p(e^{+}) + p(e^{-})] = 2 \sqrt{s}\:E(f^{+}).
\end{equation}
Some  details of the calculation have been described in
  $^{\cite{ap:fer}}$,
including the n-dimensional treatment of soft photon
contributions  which follows the methods developed in
  $^{\cite{np:bas}}$.
 The  hard photon integration has been done with SCHOONSCHIP using
tables of  integrals    $^{\cite{du:abf}}$.
 The total cross section  \st and the  integrated
 forward-backward  asymmetry without cut
  $^{\cite{du:bbf}}$ have
been determined with this technique as well as the angular
distribution  which remained unpublished so far with exclusion of
the pure QED case    $^{\cite{ap:fer}}$.
Another  set  of integration  variables allows for  a cut on the photon energy:
\begin{equation}
   \int d \Gamma = \frac{\pi^{2}}{4 s}
   \int_{-c_{m}}^{c_{m}} d\cos\Theta
   \int_{0}^{v_{m}} d v
  \int_{v_{2,min}}^{v_{2,max}} d v_{2}\frac{1}{2\pi}
   \int_{0}^{2\pi} d \varphi_{\gamma},
\label{eq:iii}
\end{equation}
\begin{equation}
  v_{2,max(min)} = \frac{1}{2} v \left[1\pm v_{m}(s')^{1/2}
   \right],
\end{equation}
\begin{equation}
  v_{m}(s) = 1 - 4m_{f}^{2}/s.
\label{eq:v23}
\end{equation}
In (~\ref{eq:iii}), $\varphi_{\gamma}$ is the azimuthal
angle of the photon in the cms and
\begin{equation}
  v_{2} = - \frac{2}{s} p(f^{+}) p(\gamma) = 1 - 2 E (f^{-})/\sqrt{s}.
\end{equation}
We should also comment on the integration boundaries for cos$\Theta$:
\begin{equation}
  c_{m} = 1 - 2 m_{e}^{2}/s.
\label{eq:ccc}
\end{equation}
The radiators for the angular  distribution contain terms behaving in
the hard photon corner like ln(1$\pm c$), (1$\pm c)^{-2}$
etc. which   become  singular and for initial state radiation
even non-integrable at  $c = \pm$1. In fact these  quantities  are
 related to $t = -2 p(e^{+}) p(f^{+}$) and $x$:
\begin{equation}
  t = \frac{x}{2} - \frac{1}{2 s} \sqrt{s^{2}-4m_{e}^{2} s }\:
                 \sqrt{x^{2}-4m_{f}^{2}s}\: \cos\Theta,
\end{equation}
\begin{equation}
  t \approx \frac{x}{2} [1-(1-2m_{e}^{2}/s) \cos\Theta].
\end{equation}
 
From (~\ref{eq:ccc}), it becomes evident that the phase space integrals
 (~\ref{eq:idg},\ref{eq:iii})
are properly regulated.  In order to get a
convolution representation for d$\sigma$/d$\cos\Theta$
 one has to perform in  (~\ref{eq:iii})
 two  integrations.  This  has been done with  SCHOONSCHIP
and REDUCE,  based on tables of integrals for hard  bremsstrahlung.
The soft photon part is the same as determined
earlier; see,  e.g.,  in  $^{\cite{cp:bkj}}$$^{\cite{np:gps}}$.
  While all the hard photon parts of
 the radiators  for the angular distribution remain
independent  of the  type of exchanged gauge boson,  this is not
the case  for the  box diagram contributions to the soft photon
part of  the QED  interference  corrections  as has been indicated
  in  the notation in (~\ref{eq:r22})
. After summing up the vertex or
box contributions  with those due to soft photons,
 the resulting $S_{A}^{a}(c,\epsilon;m,n)$ are infra-red finite but yet
 dependent on the
infinitesimal  parameter $\epsilon$  which discriminates between
soft  and  hard photons.  The  integral  over the photon energy in
 (~\ref{eq:r22})
 as  a whole is independent of $\epsilon$  as may be seen from
the formulae  of the following section. Restricting  oneself to a
semi-analytic (i.e.  containing one numerical integration)
result for d$\sigma$/d$c$, one  can take into account even cuts on the
acollinearity  and on  the fermion energies as has been
demonstrated recently  in   $^{\cite{pl:bss}}$
 where also  (~\ref{eq:iii})  was the starting point.
 
 
\vspace{1.cm}
     3. INITIAL STATE RADIATION
\vspace{1.cm}
\nopagebreak
 
In  this  chapter,  we give a systematic presentation  of
initial state corrections.
 
 
\vspace{1.cm}
     3.1. THE RADIATOR FUNCTIONS FOR THE ANGULAR DISTRIBUTION
\vspace{1.cm}
\nopagebreak
 
     There are two radiator functions for the QED inital state
     corrections,  both  of them containing well-known soft  photon
     parts  and   hard  photon  parts  which  will  deserve
some comments.
     The soft photon parts including vertex corrections are:
\begin{equation}
  S_{A}^{e}(c,\epsilon;m,n) = D_{A}(c) [ 1 + S(\epsilon,\beta_{e}) ],
\hspace{1cm} A = T,FB,
\label{eq:s31}
\end{equation}
\begin{equation}
    D_{T}(c) = 1 + c^{2},
\label{eq:d32}
\end{equation}
\begin{equation}
    D_{FB}(c) = 2c,
\label{eq:d33}
\end{equation}
where we include in  (~\ref{eq:s31}) also the Born cross section.
Further,
\begin{equation}
S(\epsilon,\beta_{e}) = \beta_{e}\left( \ln\epsilon + \frac{3}{4}\right)
+ \frac{\alpha}{\pi} Q_{e}^{2}\left(\frac{\pi^{2}}{3} - \frac{1}{2}
\right),
\label{eq:s34}
\end{equation}
\begin{equation}
   \beta_{e} = \frac{2\alpha}{\pi} Q_{e}^{2} (L_{e} - 1),
   \hspace{1cm} L_{e} = \ln \frac{s}{m_{e}^{2}}.
\end{equation}
The  hard  photon parts are symmetrised for  the  C-even  part
($A = T$) and anti-symmetrised for the C-odd part ($A = FB$):
\begin{equation}
   H_{T,FB}^{e}(v,c) = \frac{\alpha}{\pi} Q_{e}^{2}
  [ h_{T,FB}^{e}(v,c) \pm h_{T,FB}^{e}(v,-c) ] .
\label{eq:s36}
\end{equation}
As already stated,  the hard radiator parts depend on only two
variables $v, c$:
\begin{eqnarray}
   h_{T}^{e}(v,c) = \frac{z^{2}}{v^{3}}\left[ \frac{L_{c}}{\gamma^{2}}
    r_{2}\left(r_{2} - \frac{2z}{\gamma} r_{1}
    + \frac{2z^{2}}{\gamma^{2}}\right)\right. +
    (-\frac{2}{3z} r_{4} + \frac{10}{3} r_{2} - 4z) + \frac{2}{3\gamma
    z} (r_{2} r_{3}) - \nonumber \\
 -  \frac{1}{\gamma^{2}} (3r_{4} + 8 z r_{2} + \frac{26}{3} z^{2})
   + \frac{z}{\gamma^{3}} ( 8 r_{3} + \frac{44}{3} z r_{1}) -
   \frac{z^{2}}{\gamma^{4}} \left(\left.
    \frac{22}{3} r_{2} + 4 z\right)\right] ,
\end{eqnarray}
\begin{equation}
    h_{FB}^{e}(v,c) = \left.\frac{z^{2}}{v^{2}}\right[ r_{2}
    \left.\frac{L_{c}}{\gamma^{2
   }} (r_{1} - \frac{2z}{\gamma}) + \frac{2}{\gamma} r_{2}
    - \frac{4}{\gamma^{2}} r_{1} (r_{2} + z) + \frac{2z}{\gamma^{3}}
   ( 3 r_{2} + 2 z)\right].
\end{equation}
The following abbreviations are used:
\begin{equation}
    L_{c} = \ln\frac{\gamma^{2}}{z} + L_{e},
\end{equation}
\begin{equation}
    z = 1 - v, \hspace{1cm} r_{n} = 1 + z^{n},
\end{equation}
\begin{equation}
    c_{\pm} = \frac{1}{2}(1 \pm c),\hspace{1cm} \gamma = c_{+} + zc_{-}.
\label{eq:c311}
\end{equation}
As  one  should expect,  the distributions are singular for  $v$
approaching  1  (hard photon emission) and  $c$  approaching   $\pm$1
(collinear radiation). In the soft photon corner
 ($v \rightarrow \epsilon$), the
hard radiators become singular too:
\begin{equation}
    \lim_{v \rightarrow \epsilon} H_{A}^{e}(v,c) = \frac{\beta_{e}}{v}
     D_{a}(c)    + O(1), \hspace{1cm} A = T,FB.
\label{eq:lim}
\end{equation}
From  (~\ref{eq:lim})  the necessary compensation of the dependence  on
the soft photon parameter $\epsilon$  in (~\ref{eq:s34}) is evident.
Performing  the integration over the photon energy variable  $v$
leads  to analytic expressions for the initial  state
corrections  $R_{A}^{e}(c;m,n)$ to d$\sigma$/d$c$ as defined in
(~\ref{eq:d21}-\ref{eq:r22}).
These integrations are cumbersome but not too complicated. The
structure of  the  integrands $\sigma_{A}^{e,0}(s,s';m,n)
   H_{A}^{e}(v,c)$ leads  to
 integrals  containing  rational functions,  logarithms and Euler
dilogarithms. The interested reader may envisage the Fortran
program MUCUTCOS  in the package ZBIZON
   $^{\cite{zb:bar}}$  where  the
corresponding analytic  expressions for the angular distribution
 are  coded. We  only would like to comment on the origin of the
 radiative tail  within the present formalism.  For initial state
radiation, it is useful to linearise in $s'$  the resonating
function  (~\ref{eq:s10}-\ref{eq:c211}),
\begin{equation}
    \frac{1}{s' - m_{n}^{2}} \: \frac{1}{s' - m_{p}^{2*}} =
    \frac{1}{m_{p}^{2*} - m_{n}^{2}  }
\left[  \frac{1}{s' - m_{p}^{2*}} - \frac{1}{s' - m_{n}^{2}}\right],
\end{equation}
if  at least one of the interfering gauge bosons is  massive.  For
$n = p$, there arises the factor
\begin{equation}
    \frac{s}{m_{n}^{2*} - m_{n}^{2}} = -\frac{i}{2}\:\frac{M_{n}}
      {\Gamma_{n}} \: \frac{s}{M_{n}^{2}},
\label{eq:s14}
\end{equation}
setting  the  scale of tail effects, $M_{n}/\Gamma_{n}$.
The  threshold  is
defined  by  the onset of influence of the imaginary  quantity
 (~\ref{eq:s14})
onto the real type cross section.  So, it needs another
imaginary  quantity  which  is found from $ (s'- m_{n}^{2})^{-1}$.
After integration over the photon energy, functions of the following
type arise:
\begin{equation}
    \int_{\epsilon}^{\Delta} d v \frac{1}{1-v - R_{n}} =
    -\ln \frac{1-R_{n}-\epsilon}{1-R_{n}-\Delta} = -L_{R_{n}}(\Delta),
\label{eq:i315}
\end{equation}
\begin{equation}
  R_{n} = m_{n}^{2} / s.
\label{eq:r316}
\end{equation}
If  the cut-off $\Delta$ is not as infinitesimal as $\epsilon$
  (i.e.  if hard
photons are radiated),  a substantial imaginary part is
developed if $s > M_{n}^{2}$ .
 
  The photonic case, $n = p = 0$, must be treated with care.
 In order to get the exact hard bremsstrahlung correction
for pure photon exchange one must use a reduced  Born
cross  section (~\ref{eq:s10}) which  includes  an  additional
phase  space factor under the integral:
\begin{equation}
  \sigma_{A}^{0}(s',s';0,0) \Rightarrow \sqrt{1 - 4 m_{f}^{2}/s'} \:
 \: (1 + 2 m_{f}^{2}/s') \: \sigma_{A}^{0}(s',s';0,0).
\label{eq:s317}
\end{equation}
 The additional threshold factor in (~\ref{eq:s317}) influences
only  a minor part of the calculation and it is  not  too
difficult (though lengthy) to take it into account in an analytic calculation.
The  pure photonic  corrections to the angular distribution may be found
in   $^{\cite{ap:fer}}$
. The function $F_{0}$ defined there in eqs.(5), (10) corresponds
to  $R_{T}^{e}(c;0,0)$ introduced here in (~\ref{eq:r22}).  Interesting
enough,  in  the  existing  literature there  is  no  explicit
mentioning  of  the correction $R_{FB}^{e}(c;0,0)$ or of its  integral
which contributes to \afb (see below).  Since QED corrections due to photon
exchange  do not contain axial-type couplings, there  does  not
arise  any contribution of $R_{FB}^{e}(c;0,0)$ to the cross section - if
not  at  least one beam polarisation and  one  final  particle
helicity are nonzero; see (~\ref{eq:c14}-~\ref{eq:h217}).
 
 
\vspace{1.cm}
     3.2. SOFT PHOTON EXPONENTIATION
\vspace{1.cm}
\nopagebreak
 
     As  may  be  seen  from (~\ref{eq:lim}), the hard  parts  of  the
radiator  functions become singular in the soft photon  corner          of
the phase space.  Their divergent part can be combined with
the  soft photon radiator (~\ref{eq:s31},\ref{eq:s34})
 in order to get  the  lowest order
of an infinite sum over soft photon contributions.  The
treatment  can follow exactly the arguments given  in
 $^{\cite{np:gps,np:bbn}}$  for  the
total  cross section.  The result is  the  following
modification  of the initial state radiative corrections  for
the angular distribution:
\begin{equation}
   \bar{R}_{A}^{e}(c;m,n) = \int_{0}^{\Delta} d v
    \frac{\sigma_{A}^{0}(s',s';m,n)}{\sigma_{A}^{0}(s ,s ;m,n)}
   \{D_{A}(c) [ 1 + \bar{S}(\beta_{e})]
  \beta_{e} v^{\beta_{e}-1} + \bar{H}_{A}^{e}(v,c)\},
\end{equation}
\begin{equation}
 \bar{H}_{A}^{e}(v,c) = H_{A}^{e}(v,c) - \frac{\beta_{e}}{v} D_{A}(c) ,
\end{equation}
\begin{equation}
 \bar{S}(\beta_{e}) = \frac{\alpha}{\pi} Q_{e}^{2} \left[
 \frac{\pi^{2}}{3}
  - \frac{1}{2} + \frac{3}{2}(L_{e}-1)\right].
\end{equation}
     In fig. 1 the contributions of the initial state radiation
corrections  to  the differential cross section are  shown  as
functions of the scattering angle.
 
  We choose as typical energies $\sqrt{s}$ = 30 GeV (TRISTAN); 91.1 GeV
 (the Z boson mass value; LEP,SLC) and 200 GeV (tail region, LEP 200).
 The relative importance of the cross section contributions depend on three
different components; the coupling combinations $C_{A}(m,n)$ which are in
addition channel dependent, the factors (1,$\chi,|\chi|^{2}$) which have been
split away for reasons explained above, and then the Born plus QED correction
factors. Shown are the contributions for soft photon exponentiation.
 At lower energies, pure photonic corrections naturally dominate for \st
while they are zero for $\sigma_{FB}$ (if there are no polarisation phenomena).
At the Z peak and beyond in the tail region, the Z exchange dominates.
For  $|\cos\Theta|$ approaching 1 (beam directions), the cross section
rises extremely due to the collinear hard photon emission.
In fig. 2 the net Born plus initial state radiation cross section
is shown as function of the scattering angle with the photon energy cut-off
$\Delta$ as parameter. In the lower energy range the angular distribution
is nearly symmetric due to the dominance of pure QED. The same is true
at resonance, but here it is due to the smallness of the coupling
combination accompanying the nonsymmetric pure Z exchange contribution
(in case of muon production). In the tail region, a pronounced non-symmetric
angular behaviour occurs. Hard bremsstrahlung leads to a rising of the
cross section with exclusion of the resonance region where due to initial
state radiation the reduced effective energy falls below the peaking
value. A cut on the maximal photon energy reduces the cross section while
soft photon exponentiation compensates at least partly the negative
virtual plus soft photon terms.
 
\vspace{1.cm}
     3.3 INTEGRATED  CROSS SECTION AND FORWARD BACKWARD ASYMMETRY
\vspace{1.cm}
\nopagebreak

     Integrating  over the angular dependence of  the  initial
state  radiation  (~\ref{eq:s31},~\ref{eq:s36})
,  one gets the radiator functions for  \st
 $^{\cite{np:bom}}$  and \afb    $^{\cite{pl:dyb}}$.
   These are definitely different from  each
other already in the \oalf leading logarithmic approximation.
To  some extent,  this remained unclear in the comment on that
point in
 $^{\cite{pr:bbn}}$.
Nevertheless, at the Z peak the differences are
small $^{\cite{pl:jaw}}$. For more details we refer to  $^{\cite{pl:dyb}}$
and references therein.
 
  The one remaining integral over the photon
momentum can  also be calculated.  For that purpose,
one has to take care of the energy-dependence  of the
width function in
the Z boson propagator  (~\ref{eq:c211}).  Sufficiently far
away from fermion  thresholds,  approximation  (~\ref{eq:g213})
is excellent.  This fact
 allows  to apply  the Z boson transformation
 $^{\cite{pl:blr}}$
thus ensuring  that the  $s$-dependence  of  the Z
width does  not
complicate  the analytic integrations:
\begin{equation}
   \chi_{1}'(s) = \chi_{1}(s),
\end{equation}
\begin{equation}
  \chi_{1}'(s) = \frac{g_{1}^{2}}{4\pi\alpha}
  \left( 1 + i \frac{\Gamma_{Z}}
  {M_{Z}}\right)^{-1} \frac{s}{s - M_{Z}^{'2} + i M_{Z}'\Gamma_{Z}
              '},
\end{equation}
\begin{equation}
  M_{Z}' = M_{Z} / \sqrt{1 + \frac{\Gamma_{Z}^{2}}{M_{Z}^{2}}}
       \simeq M_{Z} + \Delta_{Z},
\label{eq:m323}
\end{equation}
\begin{equation}
  \Delta_{Z} = -\frac{1}{2} \frac{\Gamma_{Z}^{2}}{M_{Z}} \simeq -37 MeV,
\end{equation}
\begin{equation}
  \Gamma_{Z}' = \Gamma_{Z}/ \sqrt{ 1 + \frac{\Gamma_{Z}^{2}}{M_{Z}^{2}}
            } \simeq \Gamma_{Z},
\end{equation}
where we used M$_{Z}$ = 91.1 GeV and $\Gamma_{Z}$ = 2.6 GeV
 $^{\cite{pl:lep}}$ .
As  a result,  one gets for the total cross section (~\ref{eq:s23})  the
QED correction functions $R^{e}_{T}(m,n)$
to order  \oalf and $\bar{R}_{T}^{e}(m,n)$
with  soft photon exponentiation:
\begin{equation}
   R_{T}^{e}(m,n) = 1 + S(\epsilon,\beta_{e}) + H_{2}^{e}(m,n) +
                             H_{1}^{e}(m,n),
\end{equation}
\begin{equation}
   \bar{R}_{T}^{e}(m,n) = [ 1 + \bar{S}(\beta_{e}) ]
    H_{3}^{e}(m,n)
                           + H_{1}^{e}(m,n).
\end{equation}
The  soft  photon  corrections  $S(\epsilon,\beta_{e}$)
and $\bar{S}(\beta_{e}$)  have  been
introduced  above.  The  hard  photon  parts  consist  of  the
residual hard part $H_{1}^{e}(m,n)$ and another one, $H_{2,3}^{e}(m,n)$
, which
depends on the treatment of the soft photon corner of the hard
photon phase space integral:
\begin{equation}
  H_{i}^{e}(m,n) = g [ K_{i}(R_{m}) - K_{i}(R^{*}_{n}) ],
  \hspace{1cm} i = 1,3.
\end{equation}
Here, $g$ is a resonating kinematic factor:
\begin{equation}
 g = \frac{(1-R_{m})(1-R_{n}^{*})}{R_{m} - R_{n}^{*}}.
\end{equation}
The  threshold behaviour and further details of  the  dynamics
are contained in the $K$-functions:
\begin{equation}
  K_{1}(R) = -\frac{1}{2}\beta_{e} R [ \Delta + (1+R) I (1-R) ],
\end{equation}
\begin{equation}
   K_{2}(R) = +\frac{1}{1-R}\beta_{e}  [ R\: I(1-R) - I(0) ],
\end{equation}
\begin{equation}
  K_{3}(R) = \frac{R}{1-R} \Delta^{\beta_{e}} J\left(
 \frac{\Delta}{1-R}\right),
\end{equation}
\begin{equation}
 I(z) = - \int_{\epsilon}^{\Delta} dv \frac{1}{v-z} = -\ln \frac
    {\Delta-z}{\epsilon-z},
\end{equation}
\begin{equation}
   J(\alpha) = _{2}\!F_{1}(1,\beta_{e},1+\beta_{e},\alpha) = \beta_{e}
     \int_{0}^{1} dv \frac{v^{\beta_{e}-1}}{1-\alpha v}.
\end{equation}
     The above  definitions are valid if at least  one of  the
$R_{m},R_{n}$ are non-zero.
 
 Although the limit of both masses $M_{m}, M_{n}$ vanishing formally
exists, it differs from the correct form of $R_{T}^{e}(0,0)$ by a constant.
This is due to the occurence of the double pole $(1-v)^{-2}$ from
$\sigma^{e}_{T}(s',s';0,0)$ under the integral. Such a double pole
behaviour cannot
be calculated as continuous limit of the two single poles arising
in the massive case: 
\begin{eqnarray}
 \bar{R}_{T}^{e} (0,0) =  \left( \begin{array}{c}
                           \Delta^{\beta_{e}} \\ 1 \end{array} \right)
                                      \left\{ \begin{array}{c}
                          J(\Delta) \\ 1 \end{array} \right\}
         [ 1 + \bar{S}(\beta_{e}) + \delta_{2}^{V+S} ]
      + \frac{2\alpha}{\pi} Q_{e}^{2} (L_{e}-1)
\left( \mp \frac{1}{2} \right)
                \ln \frac{\Delta}{1-\Delta},
\end{eqnarray}
where the upper, barred (lower, unbarred) case is with (without) soft
photon exponentiation.
Further, if the cut-off $\Delta$ is removed,
an additional finite error would occur due to lacking phase space
factors (~\ref{eq:s317}). 
We quote here the corresponding correction without
$^{\cite{np:kum,du:bbf}}$ and with soft photon exponentiation:
\begin{eqnarray*}
 \bar{R}_{T}^{e} (0,0) =  
                                      \left\{ \begin{array}{c}
                          J(\rho) \\ 1 \end{array} \right\}
         [ 1 + \bar{S}(\beta_{e}) + \delta_{2}^{V+S} ]
      + \frac{2\alpha}{\pi} Q_{e}^{2}(L_{e}-1) 
\left[ \left(\mp \frac{1}{2}\right) \ln \frac{s}{m_{f}^{2}} + \frac{2}{3}
\left( \begin{array}{c} 1 \\-2 \end{array} \right) \right],
\end{eqnarray*}
\begin{eqnarray*}
\rho = 1 - 4m_{f}^{2}/s.  
\end{eqnarray*}
The logarithmic final state mass
dependence regularises the massless photon propagator for emission
of very hard photons and the corresponding singularity of ln(1-$\Delta$)
in the limit $\Delta \rightarrow 1$.
For applications,  we  add to the soft photon correction
$\bar{S}(\beta_{e}$) the
next to leading order logarithmic correction $\delta_{2}^{V+S}$
       exactly  as obtained  in
   $^{\cite{np:bbn,yr:fab}}$.
For $H_{1}^{e}(1,1)$ which is the dominating  hard
photon contribution at LEP energies, we also add terms arising
from  an  explicit integration with  the  corresponding  hard
photon  terms in $\delta_{2}^{H}$  as quoted in
  $^{\cite{np:bbn,yr:fab}}$.
This ensures  the
necessary  high  accuracy  of the order $O(0.1\%)$  but  is  not
described here.
 
     Due  to  the  occurence of  logarithms  in  the hard radiator
function for $\sigma_{FB}$ (see $^{\cite{pl:dyb}}$ for details),
the corresponding integrated expressions have a slightly
more complicated structure than those for \st. For the photon
exchange contribution,
\begin{eqnarray}
 \bar{R}_{FB}^{e} (0,0) =   \bar{R}_{T}^{e} (0,0)
+ \frac{2\alpha}{\pi} Q_{e}^{2} \left\{ \frac{1}{2}
 (L_{e}-1) \ln \frac{\Delta}{1-\Delta} 
                                           \right. 
     \nonumber \\
+ 4d + (L_{e}-1)[L_{1} + L_{2} + 2d ] + 2\left.\frac{1-\Delta}
               {2-\Delta} L_{1}
  + 4 d L_{2} +  l_{4} + l_{3}\right\}.
\end{eqnarray}
The following abbreviations are used:
\begin{equation}
 d = \frac{1}{2} \frac{\Delta}{2-\Delta},
\end{equation}
\begin{equation}
   L_{1} = \ln(1-\Delta),
\end{equation}
\begin{equation}
  L_{2} = \ln\left(1-\frac{\Delta}{2}\right),
\end{equation}
\begin{equation}
  l_{3} =  Li_{2}(\Delta) - 2 Li_{2}\left(
  \frac{\Delta}{2}\right),
\end{equation}
\begin{equation}
 l_{4} = Li_{2}\left(\frac{1}{2}\right)
 - Li_{2}\left(\frac{1}{2-\Delta}\right) -
   \ln 2 L_{2} + \frac{1}{2} L_{2}^{2},
\end{equation}
\begin{equation}
  Li_{2}(z) = -\int_{0}^{1}\frac{dx}{x} \ln(1-xz).
\end{equation}
The initial state correction to $\sigma_{FB}$ due to the $\gamma$Z
interference is:
\begin{eqnarray}
 \bar{R}_{FB}^{e} (1,0) =  \left( \begin{array}{c}
                           \Delta^{\beta_{e}} \\ 1 \end{array} \right)
                                      \left\{ \begin{array}{c}
                          J[\Delta/(1-R)] \\ 1 \end{array} \right\}
         [ 1 + \bar{S}(\beta_{e}) + \delta_{2}^{V+S} ]  \nonumber \\
       + \frac{2\alpha}{\pi} Q_{e}^{2}  \left\{
                                      \left( \begin{array}{c}
                   0 \\ 1 \end{array} \right) \right.
  (L_{e}-1)[\ln\Delta + L_{R}(\Delta)]     \nonumber \\
    + 2(L_{e}-1)  [ d\:r - \frac{1}{2}L_{R}(\Delta)] + 4\:d\:r
    + 2 r \frac{1-\Delta}{2-\Delta}L_{1} \nonumber \\
 + \left[\frac{(L_{e}-1)}{1+R}(1+3R)
 + 4\:d \right]r L_{2} \nonumber \\
+(1-2R)l_{4} + l_{3} +
2R\left.\frac{1+R^{2}}{(1+R)^{2}}D_{3}\right\}.
\end{eqnarray}
The resonance logarithm $L_{R}(\Delta)$ as defined in  (~\ref{eq:i315})
and the additional abbreviations depend  on the complex Z boson mass
parameter:
\begin{equation}
   D_{3} = 2D_{2} - D_{1} + l_{4} + (L_{e}-1-2\ln2)L_{R}(\Delta),
\end{equation}
\begin{equation}
  D_{2} = Li_{2}\left(\frac{2}{1+R}\right)
  - Li_{2}\left(\frac{2-\Delta}{1+R}\right)
  + \ln2\ln(-r) - (L_{2}+\ln 2)\ln\left(1-\frac{2-\Delta}{1+R}\right),
\label{eq:d74}
\end{equation}
\begin{equation}
  D_{1} = Li_{2}\left(\frac{1}{R}\right)
  - Li_{2}\left(\frac{1-\Delta}{R}\right)
   - L_{1}\ln\left(1-\frac{1-\Delta}{R}\right),
\label{eq:d75}
\end{equation}
\begin{equation}
   r = \frac{1-R}{1+R}.
\end{equation}
The third correction to $\sigma_{FB}$ is due to Z exchange:
\begin{eqnarray}
 \bar{R}_{FB}^{e} (1,1) =  \left( \begin{array}{c}
                           \Delta^{\beta_{e}} \\ 1 \end{array} \right)
                                      \left\{ \begin{array}{c}
           Im[R(1-R^*)J(\frac{\Delta}{1-R})]       / Im(R)
          \\ 1 \end{array} \right\}
         [ 1 + \bar{S}(\beta_{e}) + \delta_{2}^{V+S} ]  \nonumber \\
      + \frac{2\alpha}{\pi} Q_{e}^{2} \left\{ \left( \begin{array}{c}
                           0 \\1 \end{array} \right)\right.
               (L_{e}-1) [ \ln\Delta + t\:L_{R}(\Delta)]
                                                       \nonumber \\
 - t\:(L_{e}-1)L_{R}(\Delta)
      + 2|r|^{2}d(L_{e}+1)
      + 2R \frac{1+R^{2}}{(1+R)^{2}}tD_{3}
   + 2 |r|^{2}\frac{1-\Delta}{2-\Delta} L_{1} + 4d |r|^{2} L_{2}
                                                       \nonumber \\
 + (L_{e}-1)L_{2}[8\frac{R^{2}-1}{|1+R|^{4}} + 4\frac{1-6R}{|1+R|^{2}}
   + 5]
 +  (-1+2|1-R|^{2})l_{4} + l_{3}\},
\end{eqnarray}
\begin{equation}
   t = R\frac{2(1-R^{*})}{R-R^{*}}.
\end{equation}
The C-odd initial state corrections behave very similar to those for
\st. They have also the logarithmic electron mass singularity $L_{e}$.
Both Z exchange corrections $R_{A}^{e}(1,1)$ develop a radiative tail
beyond the resonace. This is due to their structure as already discussed
for the angular distribution; see (~\ref{eq:i315}):
\begin{equation}
 R_{A}^{e}(1,1) = f_{A}^{1}(R) + f_{A}^{2}(R)tL_{R}(\Delta)(L_{e}-1).
\end{equation}
Because of their complexity, we quote here the functions $R_{A}^{e}(1,0)$
and $R_{A}^{e}(1,1)$ for $\Delta$ = 1
 taken at  $R$ = 1, i.e. at resonance
 (we leave out here the coupling constants in the definition of $\chi$
(~\ref{eq:c211})):
\begin{equation}
  |\chi|^{2}Re \:R_{A}^{e}(1,1)|_{R=1} \sim \frac{1}{\rho^{2}}
      [1+\frac{\alpha}{\pi}Q_{e}^{2}(h_{1}^{e}+\rho h_{2}^{e} + \rho^{2}
      h_{A}^{e})],
  \hspace{.5cm}     A = T,FB,
\end{equation}
\begin{equation}
    \rho = \Gamma_{Z}/M_{Z},
\end{equation}
\begin{equation}
  h_{1}^{e} = -\frac{1}{2} + 2 Li_{2}(1) + (L_{e}-1)[\frac{3}{2}+2\ln \rho],
\end{equation}
\begin{equation}
 h_{2}^{e} = -2\pi (L_{e}-1),
\end{equation}
\begin{equation}
 h_{T}^{e} = (L_{e}-1)(2-3\ln \rho),
\end{equation}
\begin{equation}
 h_{FB}^{e} = \frac{9}{4} - \frac{7}{2}\ln 2 + 4\ln^{2}2 - \frac{3}{4}
  Li_{2}(1) - \frac{1}{2}\ln\rho + (L_{e}-1)(-\frac{3}{2}\ln 2 + \frac{7}{2}
 - \frac{5}{2}\ln\rho).
\end{equation}
From these \oalf expressions it is easy to derive the corresponding
ones for soft photon exponentiation. The most interesting feature
is connected  with hard photons - or better, with their absence at resonance.
The initial state emission of a hard photon for $s = M_{Z}^{2}$ leads
to a largely reduced effective energy $s'$ and thus a non-resonant behaviour,
i.e. a much reduced cross section. As a consequence of the resulting
soft photon dominance, C-even and C-odd observables behave similar and
consequently the leading order coefficients $h_{1}^{e},h_{2}^{e}$ are
equal for them at resonance. This has been observed numerically first in
$^{\cite{pl:jaw}}$ and explained in $^{\cite{pl:dyb}}$.
The $\gamma$Z interference corrections at the peak position are
exclusively due to the imaginary parts of $R_{A}^{e}(1,0)$ since $\chi$
becomes imaginary for $R$ = 1:
\begin{equation}
 Re[\chi R_{A}^{e}(1,0)] \sim \frac{1}{\rho}\{\delta_{A,T} +
    \frac{\alpha}{\pi}Q_{e}^{2}[\pi(L_{e}-1) + \rho g_{A}^{0} +
    \rho^{2} g_{A}^{1}]\},
\end{equation}
\begin{equation}
 g_{T}^{0} = 2(L_{e}-1)(\ln\rho - \frac{1}{2}), \hspace{1cm}
 g_{T}^{1} = -\frac{\pi}{2}(L_{e}-1),
\end{equation}
\begin{equation}
 g_{FB}^{0} = 2(L_{e}-1)(\ln\rho - \frac{3}{2} + \ln 2) +
 [-2 + 2\ln 2 - 4\ln^{2}2 + Li_{2}(1)],
\end{equation}
\begin{equation}
 g_{FB}^{1} = -\frac{\pi}{4} L_{e}.
\end{equation}
The Born contribution to $\sigma_{FB}$ vanishes at the peak. Nevertheless,
 the leading corrections to \st and $\sigma_{FB}$ are equal again.
Integrated cross section and asymmetry as functions of energy and the
photon energy cut-off $\Delta$ have been studied in detail in
$^{\cite{du:bbf,ph:bis}}$. We will come back to that point in chapters
5 and 6.
 
 
\vspace{1.cm}
       4. INITIAL-FINAL STATE INTERFERENCE RADIATION
\vspace{1.cm}
\nopagebreak
 
     The  initial-final state interference corrections have an
interesting  property:  only those which are diagonal  in  the
arguments  indicating  the exchanged vector bosons  are  independent
observables.  The other interference contributions may
be determined by the following simple relation:
\begin{equation}
 R_{A}^{i}(c;m,n) = \frac{1}{2}[R_{A}^{i}(c;m,m) + R_{A}^{i}(c;n,n)^{*}]
\label{eq:r41}
\end{equation}
In order to get (~\ref{eq:r41}), it is essential to separate the
reduced  Born factor (~\ref{eq:s09},~\ref{eq:s10})
from the QED contents.  The  validity  of
(~\ref{eq:r41})  may be seen immediately from (~\ref{eq:r22},~\ref
{eq:s10},~\ref{eq:c211}).
  An explicit proof has been given in $^{\cite{zp:bbc}}$ .
 
 
\vspace{1.cm}
       4.1. THE RADIATOR FUNCTIONS FOR THE ANGULAR DISTRIBUTIONS
\vspace{1.cm}
\nopagebreak
 
     For interference corrections, the soft photon part has to
include besides the soft photon emission terms
$S_{A}(c,\epsilon,\lambda)$ also
the  contributions  $B_{A}(c,\lambda;m,n)$ originating
from  photon  - Z
boson  and photon - photon box diagrams.  Their sum is infrared
finite  (and independent of the infrared cut-off $\lambda$) while  the
infinitesimal  soft photon cut-off parameter $\epsilon$ disappears only
after integrating  over the photon energy variable $v$:
\begin{equation}
 S_{A}^{i}(c,\epsilon;m,n) = \frac{\alpha}{\pi}Q_{e}Q_{f}[S_{A}(c,
 \epsilon,\lambda) + B_{A}(c,\lambda;m,n)],
\label{eq:s80}
\end{equation}
\begin{equation}
  S_{A}(c,\epsilon,\lambda) = 2 D_{\bar{A}}(c)[2\ln\frac{\epsilon}{
  \lambda}\ln\frac{c_{-}}{c_{+}} + Li_{2}(c_{+}) - Li_{2}(c_{-})
 -\frac{1}{2}(\ln^{2}c_{+} - \ln^{2}c_{-})].
\end{equation}
Here, $D_{\bar{A}}(c)$ is defined in (~\ref{eq:d32}-~\ref{eq:d33})
with $\bar{A} = T(FB)$ if $A = FB(T)$,
and $c_{\pm}$ in (~\ref{eq:c311}).
The  box diagram contributions   $^{\cite{np:bcf}}$
 are dependent on the  type
of  the  gauge bosons exchanged.  We  write  them  in
(anti-) symmetrised form:
\begin{equation}
 B_{T,FB}(c,\lambda;m,n) = b(c,\lambda;m,n) \pm b(-c,\lambda;m,n).
\end{equation}
The two different box functions are:
\begin{equation}
 b(c,\lambda;0,0) = -2c_{+}^{2}\ln\frac{c_{+}}{c_{-}}[4\ln\frac{2E}{
     \lambda} - 2\pi i ] + \ln c_{-}[-2c_{+} + c(\ln c_{-} + 2\pi i)]
         - 2\pi i c_{+},
\end{equation}
\begin{eqnarray}
 b(c,\lambda;n,n) = -4c_{+}(1-R_{n})\{\ln\frac{c_{-}}{R_{n}}
 - (1-R_{n})L_{n} +
   \nonumber \\
 \frac{1}{c_{+}}(1-R_{n} - 2c_{+})[l(1) - l(c_{-}) -L_{n}\ln c_{-}]\}
   \nonumber \\
  - 2c_{+}^{2}\{[2\ln\frac{2E}{\lambda} + 4L_{n} + \ln(c_{+}c_{-})]
   \ln\frac{c_{+}}{c^{-}} + 2l(c_{+}) - 2l(c_{-})\},\hspace{1cm}
     n \neq 0.
\end{eqnarray}
The following abbreviations are used:
\begin{equation}
 l(a) = Li_{2}(1 - aR_{n}^{-1}),
\end{equation}
\begin{equation}
   L_{n} = \ln(1 - R_{n}^{-1}) \equiv L_{R_{n}}(1),
\end{equation}
and $R_{n}$  is defined in (~\ref{eq:r316}),
$L_{R}(\Delta)$ in (~\ref{eq:i315}).
 
The  hard  radiator parts are independent of the  gauge  boson
exchanged:
\begin{equation}
   H_{T,FB}^{i}(v,c) = \frac{\alpha}{\pi} Q_{e}Q_{f}
        [h_{T,FB}^{i}(v,c) \pm h_{T,FB}^{i}(v,-c)].
\end{equation}
While the box terms for $A = FB$ and $A = T$ could be expressed by
one and the same function, this is not the case here:
\begin{eqnarray}
\lefteqn{h_{T}^{i}(v,c) =} \nonumber \\
& 2\left.c_{+}\right\{\left[\frac{4}{v}\right. - 3
 - z(2+z)\left]\ln\frac{c_{-}}{c_{+}} - (1+z)^{2}\ln
 \frac{c_{-}+zc_{+}}{c_{+}+zc_{-}}\right\}
 \nonumber \\
& + 2\left[-\frac{4}{v}+4+z(2+\ln z)\right]
  + \frac{4}{\gamma}\left[\frac{2}{v}
  -2-z(1+z)\right],
\end{eqnarray}
\begin{eqnarray}
\lefteqn{ h_{FB}^{i}(v,c) =} \nonumber \\
& 2(1+c^{2})\ln c_{-}(\frac{2}{v} - 1 - z - z^{2})
\nonumber  \\
&    + 4c_{+}[-\frac{4}{v} + 4 + 2z - z(1-z)\ln z] +\frac{4}{\gamma}
   (-\frac{2}{v^{2}}+ \frac{5}{v}-3-z)
\nonumber \\
& +\frac{2}{\gamma^{2}}(-\frac{2}{v^{2}}+\frac{6}{v}-4-2z-z^{2})
   +2(z^{2}-1)c_{+}\ln c_{+}c_{-} +
\nonumber  \\
&   2[(1-z+z^{2})+c(1-z^{2})+c^{2}(1+z+z^{2})]\ln\gamma.
\end{eqnarray}
It is easy to check that in the soft photon limit ($v \rightarrow$ 0) the
functions  $h_{A}^{i}(v,c)$ behave
such that they  compensate  after
integration over the photon momentum the dependence of the box
contributions on ln$\epsilon$.
In Fig. 3, the interference contributions are shown as functions of the
scattering angle for three different energies.
Compared to initial state radiation,  the interference corrections
are small.  They do not contain fermion mass
singularities  though kinematic singularities at cos$\Theta = \pm$ 1
 occur.  In contrast to initial state radiation,  these are
integrable  so that formally the full angular range in
(~\ref{eq:idg},~\ref{eq:iii}) may be used.
Of  course,  the interference contributions depend
on the hard photon cut-off $\Delta$.
 For more severe cuts, they grow up. This is
due to the fact that for the totally inclusive
problem ($\Delta = v_{m} \sim 1$) there exists a fine-tuned
cancellation of box and  bremsstrahlung
contributions (for more details see next
section)  which becomes  more  and more disbalanced
if the phase space of  the bremsstrahlung integral
becomes more and more
restricted.  For very tight cut values, $\Delta \ll$ 1,
the cross section to order \oalf starts to diverge and may become even
negative.
 
This can be seen from the leading soft photon contribution
$R_{A}^{i,soft}(c;m,n)$. After integration over the photon energy,
the dependence of $S_{A}(c,\epsilon,\lambda)$ in (~\ref{eq:s80})
on $\epsilon$ drops out. It is replaced by the following terms
arising from the soft photon corner of the phase space integral:
\begin{equation}
 R_{A}^{i,soft}(c;m,n) = 2D_{\bar{A}}(c)\beta_{i}\ln \Delta,
\label{eq:a1}
\end{equation}
\begin{equation}
 \beta_{i} = 2\frac{\alpha}{\pi}Q_{e}Q_{f}\ln\frac{1-c}{1+c}.
\label{eq:a2}
\end{equation}
The complete interference contributions to the differential cross section
may be found in the code MUCUTCOS  $^{\cite{zb:bar}}$. In analogy to
initial state radiation, the logarithmic dependence on the cut-off $\Delta$ in
(~\ref{eq:a1}) has to be cured by an adequate soft photon resummation which
should lead to a replacement of
\begin{equation}
\int_{0}^{\Delta}dv\;\sigma_{A}^{i,0}(s,s';m,n)[\delta(v)\beta_{i}\ln\epsilon
+ \Theta(v-\epsilon)\frac{\beta_{i}}{v}]
\label{eq:a3}
\end{equation}
by some function of the following type:
\begin{equation}
 \int_{0}^{\Delta}dv\;\sigma_{A}^{i,0}(s,s';m,n)\beta_{i}v^{\beta_{i}-1}.
\label{eq:a4}
\end{equation}
In order to get really a smooth, well-defined behaviour of (~\ref{eq:a4}), the
exponent therein has to be larger than -1, i.e. $\beta_{i}$ should be positive
definite.
This is not the case in (~\ref{eq:a4}) as may be seen from (~\ref{eq:a2}).
A possible way out is the exponentiation of interference radiation together
with initial and final state radiation as is dictated by soft photon theorems
(see, e.g., $^{\cite{np:gps}}$). A collection of the relevant terms
leads to the following order \oalf expressions after integration.
(One should have in mind that $s' = s$ for the soft photon case):
\begin{equation}
\frac{d\sigma^{soft}}{dc} = \sum_{A,m,n}\sigma_{A}^{0}(s,s;m,n)D_{A}(c)
\\ \nonumber
 \{1+\frac{2\alpha}{\pi}[Q_{e}^{2} \ln (\frac{s}{m_{e}^{2}}-1)
+2Q_{e}Q_{f}\ln \frac{1-c}{1+c}
+Q_{f}^{2}(\ln\frac{s}{m_{f}^{2}}-1)]\ln \Delta \}.
\label{eq:a5}
\end{equation}
 The final
state expressions have been taken from sect. 5. Again, the correction
is not positive definite due to the expression in square brackets wich is
a possible candidate to replace the exponent in (~\ref{eq:a4}). This is a
consequence of too crude approximations. The soft photon contribution as a
whole must be positive for any parameter combination since it is a complete
module squared. In fact:
\begin{equation}
 d\sigma^{soft} = \sum_{A,m,n}\sigma_{A}^{0}(s,s;m,n)D_{A}(c)
(1+\frac{\alpha}{\pi}\chi^{2}),
\label{eq:a6}
\end{equation}
\begin{equation}
\chi = Q_{e}\left(\frac{k_{1}}{k_{1}p} - \frac{k_{2}}{k_{2}p}\right)
       + Q_{f}\left(\frac{p_{1}}{p_{1}p} - \frac{p_{2}}{p_{2}p}\right).
\label{eq:a7}
\end{equation}
The four momenta used here are defined in (~\ref{eq:cross}).
After a little algebra, one gets
\begin{eqnarray}
\lefteqn{ \chi^{2} =} \nonumber \\
& Q_{e}^{2}\left[\frac{s-2m_{e}^{2}}{(-k_{1}p)(-k_{2}p)} -
 \frac{m_{e}^{2}}{(-k_{1}p)^{2}} - \frac{m_{e}^{2}}{(-k_{2}p)^{2}}\right]
+ Q_{f}^{2}\left[\frac{s-2m_{f}^{2}}{(-p_{1}p)(-p_{2}p)}
- \frac{m_{f}^{2}}{(-p_{1}p)^{2}} - \frac{m_{f}^{2}}{(-p_{2}p)^{2}}\right]
\nonumber \\
& + Q_{e}Q_{f}\left\{t_{+}\left[\frac{1}{(-k_{2}p)(-p_{1}p)}
+ \frac{1}{(-k_{1}p)(-p_{2}p)}\right]
- t_{-}\left[\frac{1}{(-k_{1}p)(-p_{1}p)}
+ \frac{1}{(-k_{2}p)(-p_{2}p)}\right]\right\},
\label{eq:a8}
\end{eqnarray}
\begin{equation}
t_{+(-)} = -2p_{1(2)}k_{2} = -2p_{2(1)}k_{1} = \frac{s}{2}\pm \frac{c}{2}
\sqrt{s-4m_{e}^{2}}\sqrt{s-4m_{f}^{2}}.
\label{eq:a9}
\end{equation}
After integration over the soft photon corner of the n dimensional photon
momentum phase space $^{\cite{np:bas}}$, one gets for the
corrections which are relevant for soft photon exponentiation:
\begin{eqnarray}
 d\sigma^{soft} = \sum_{A,m,n}\sigma_{A}^{0}(s,s;m,n)D_{A}(c)
\left\{1+\frac{2\alpha}{\pi}Q_{e}^{2}\right(\frac{s-2m_{e}^{2}}{\sigma_{e}}
\ln\left.\frac{s+\sigma_{e}}{s-\sigma_{e}}-1\right)
\nonumber \\
+ \left.\frac{2\alpha}{\pi}Q_{f}^{2}\right(\frac{s-2m_{f}^{2}}{\sigma_{f}}
\ln\left.\frac{s+\sigma_{f}}{s-\sigma_{f}}-1\right)
\nonumber \\
+ \frac{2\alpha}{\pi}Q_{e}Q_{f}
\left[\frac{\sigma_{c}^{-}}{R^{-}}
\ln\right.\frac{\sigma_{c}^{-}+R^{-}}{\sigma_{c}^{-}-R^{-}}
-\frac{\sigma_{c}^{+}}{R^{+}}\left.\left.
\ln\frac{\sigma_{c}^{+}+R^{+}}{\sigma_{c}^{+}-R^{+}}\right]\right\},
\label{eq:a10}
\end{eqnarray}
\begin{equation}
 \sigma_{e(f)} = (s^{2}-4m_{e(f)}^{2}s)^{1/2},
\label{eq:a11}
\end{equation}
\begin{equation}
 \sigma^{\pm}_{c} = s^{2}\pm c\sigma_{e}\sigma_{f},
\label{eq:a12}
\end{equation}
\begin{equation}
 R^{\pm} = (\sigma_{c}^{\pm 2}-16m_{e}^{2}m_{f}^{2}s^{2})^{1/2}.
\label{eq:a13}
\end{equation}
In the ultra-relativistic limit of (~\ref{eq:a10}), one gets (~\ref{eq:a5}).
While (~\ref{eq:a10}) as a whole is positive definite, this property
is lost in (~\ref{eq:a5}) as a result of the approximations.
 
Since we do not intend in our applications to apply such stringent
cuts as to make inevitable the adequate exponentiation of initial
final interference contributions we would like to remain at this point of
clarification.
 
In a recent series of papers, there have been presented quite interesting
results on soft photon exponentiation $^{\cite{pl:agr}}$ including
the interference $^{\cite{br:gre}}$. Tracing back also to earlier work on
that subject $^{\cite{np:gps}}$, the problem of exponentiation of
soft photon interference radiation has been solved there in the
ultra-relativistic limit. From the above discussion we conclude that this
is adequate for values of $cos\Theta$ not too close to 1, i.e. assuming some
realistic acceptance cut for the experimental set-up. Further, the
authors make  the implicit assumption that one
may use the radiators for \st also
for the differential cross section. This is not the case in general as shown
in this article and in $^{\cite{pl:dyb}}$. Nevertheless, the radiators
for \st can be used as an approximation. Further, we see no
problem in combining the hard radiator parts presented here for the
differential cross section with the result of $^{\cite{br:gre}}$ for
the refined treatment of soft photon exponentiation.
 
\vspace{1.cm}
       4.2. INTEGRATED CROSS SECTION AND FORWARD BACKWARD ASYMMETRY
\vspace{1.cm}
\nopagebreak
 
     The  integrated interference contributions  are  composed
of box  corrections   $B_{A}(\lambda,n,n)$   and   of   bremsstrahlung
contributions $b_{A}(\lambda,n,n)$:
\begin{equation}
   R_{A}^{i}(n,n) =   \frac{\alpha}{\pi}Q_{e}Q_{f}
 [B_{A}(\lambda;n,n) + b_{A}(\lambda;n,n)].
\end{equation}
For the total cross section one gets:
\begin{equation}
 B_{T}(\lambda;0,0) = 6\ln\frac{2E}{\lambda} - \frac{9}{2},
\end{equation}
\begin{equation}
 b_{T}(\lambda;0,0) = -6\ln\frac{2E\Delta}{\lambda} + 3(2+\Delta),
\end{equation}
\begin{equation}
 B_{T}(\lambda;1,1) = 6\ln\frac{2E}{\lambda} - 9 + 3R[1+(1+R)
  L_{R}(1)] + 3L_{Z},
\end{equation}
\begin{equation}
 b_{T}(\lambda;1,1) = -6\ln\frac{2E\Delta}{\lambda} + 6 +
  3[\Delta(1-R) - R(1+R)L_{R}(\Delta)],
\end{equation}
\begin{equation}
  L_{Z} = \ln (s/M_{Z}'^{2}),
\end{equation}
where $L_{R}(\Delta)$ is defined in (~\ref{eq:i315}
) and $M'$ in (~\ref{eq:m323}).
Without cut ($\Delta$ = 1), the resulting corrections to \st
   are quite small
$^{\cite{du:bbf,un:jks}}$:
\begin{equation}
 R_{T}^{i}(0,0) = \frac{\alpha}{\pi}Q_{e}Q_{f}\frac{9}{2},
\end{equation}
\begin{equation}
 R_{T}^{i}(1,1) = \frac{\alpha}{\pi}Q_{e}Q_{f}3L_{Z}.
\label{eq:r419}
\end{equation}
The interference QED corrections due to photon exchange to the
antisymmetric cross section part $\sigma_{FB}$ are:
\begin{equation}
  B_{FB}^{i}(\lambda;0,0) = (1+8\ln2)\ln\frac{2E}{\lambda}
  -\frac{3}{4}(1+6\ln2+\ln^{2}2) + \frac{1}{2}i\pi(2-5\ln2),
\end{equation}
\begin{eqnarray}
  b_{FB}^{i}(\lambda;0,0) = -(1+8\ln2)\ln\frac{2E\Delta}{\lambda}
  +\frac{1}{4}(2-5\Delta) + \frac{1}{4}(4+26\Delta-5\Delta^{2})\ln2
   \nonumber \\
  + \frac{1}{4}(-44+26\Delta-5\Delta^{2})L_{2} + \frac{1}{4}
   (15-20\Delta+5\Delta^{2})L_{1} \nonumber \\
   + \frac{3}{4}Li_{2}(1) + 4\ln^{2}2 - 4Li_{2}(\Delta) + 8Li_{2}
\left(\frac{\Delta}{2}\right) -\frac{5}{2}Li_{2}(\Delta-1).
\end{eqnarray}
Their sum is a simple constant for $\Delta$ = 1
$^{\cite{ap:fer}}$:
\begin{equation}
 R_{FB}^{i}(0,0) = \frac{\alpha}{\pi}Q_{e}Q_{f}(4.572-i2.302).
\end{equation}
The Z exchange corrections are much more involved. For $\Delta$ = 1,
they may be found in $^{\cite{du:bbf}}$; with $\Delta\leq$ 1:
\begin{eqnarray}
B_{FB}^{i}(\lambda;1,1) = (1+8\ln 2)\ln \frac{2E}{\lambda} - \frac{3}{2}
 + R - (9-4R-4R^{2})\ln 2 \nonumber \\
 - 2\ln^{2}2 + \frac{1}{2}(5-4R)L_{Z} - \frac{1}{2}[4-9R+3R^{2}+
 2(-5+3R-6R^{2})\ln 2] L_{R}(1) \nonumber \\
 -(1-3R+6R^{2}-8R^{3})\left[Li_{2}(1-\right.\frac{1}{2R})
-Li_{2}(1-\left.\frac{1}{R})\right]
  - 4R^{3}\left[Li_{2}(1)-Li_{2}(1-\frac{1}{R})\right],
\end{eqnarray}
\begin{eqnarray}
  b_{FB}^{i}(\lambda;1,1) = -(1+8\ln 2)\ln\frac{2E\Delta}{\lambda} +
   \frac{1}{4}(2-5\Delta+5R\Delta) \nonumber \\
+   \frac{1}{4}(4-16R\Delta+5R\Delta^{2}-10R^{2}\Delta
   +26\Delta-5\Delta^{2})\ln2  +
 4\ln^{2}2 +2Li_{2}(1) \nonumber \\
 +  \frac{5}{4}(3-R+2R\Delta-R\Delta^{2}-2R^{2}+2R^{2}\Delta-4\Delta
   +\Delta^{2})L_{1} \nonumber \\
 + (-5+3R-4R\Delta+\frac{5}{4}R\Delta^{2}+5R^{2}-\frac{5}{2}R^{2}\Delta
 -\frac{6}{1+R}+\frac{13}{2}\Delta-\frac{5}{4}\Delta^{2})L_{2}
                   \nonumber \\
 + R(-5+\frac{6}{1+R}+R)L_{R}(\Delta) - 4Li_{2}(\Delta)+8Li_{2}(\frac{
                \Delta}{2})
  \nonumber \\
  + \frac{R}{2}(3+5R^{2})D_{1} - \frac{1}{2}(5+3R+3R^{2}+5R^{3})D_{2},
\end{eqnarray}
with $D_{1,2}$ defined in (~\ref{eq:d74}-~\ref{eq:d75}). Again, the Z boson
parameters have to be understood in primed quantities.
A generalization of the initial final interference QED corrections
to  the  case of several massive gauge bosons $^{\cite{pl:lrs}}$ is
trivial due to relation (~\ref{eq:r41}).
     At  the  Z  peak,  a fine-tuned cancellation of  box  and
bremsstrahlung  terms   occurs.  Therefore,  the
resulting Z exchange interference contributions $R_{T}^{i}(1,1)$ and $R_{FB}
^{i}(1,1)$ become small there. This has been noticed from numerical
results  in  $^{\cite{cp:bkj,du:bbf,pl:jaw}}$
and is known also from earlier investigations of J/$\psi$ physics
$^{\cite{jpsi}}$.
For $R_{T}^{i}(1,1)$ it is  evident  from (~\ref{eq:r419}).
A Taylor expansion around $s = M_{Z}^{2}$ yields:
\begin{equation}
 R_{T}^{i}(1,1) =\frac{\alpha}{\pi}Q_{e}Q_{f}
[6\ln\frac{R-1+\Delta}{\Delta} + 9(R-1)\ln (R-1+\Delta)
 + O(R-1)],
\label{eq:r102}
\end{equation}
\begin{eqnarray}
 R_{FB}^{i}(1,1) =\frac{\alpha}{\pi}Q_{e}Q_{f}\{
  [1+8\ln (2-\Delta)-4\ln (1-\Delta)]
\ln\frac{R-1+\Delta}{\Delta}
 - 3(\ln 2)(R-1)\ln (R-1)      \nonumber \\
 + \frac{3}{2} [1+8\ln (2-\Delta)-6\ln (1-\Delta)](R-1)\ln (R-1+\Delta)
 + 4(R-1)\ln (1-\Delta) + O(R-1)\}.
\label{eq:r103}
\end{eqnarray}
For $\Delta$ approaching 1 and using in the peak region
ln $R = R-1 + O[(R-1)^{2}]$, one gets
\begin{equation}
 R_{T}^{i}(1,1;\Delta=1) = \frac{\alpha}{\pi}Q_{e}Q_{f}
\{-3(R-1) + O[(R-1)^{2}]\},
\end{equation}
\begin{eqnarray}
 R_{FB}^{i}(1,1;\Delta=1) = \frac{\alpha}{\pi}Q_{e}Q_{f}\{
 -3(R-1) (\ln 2)\ln (R-1) \nonumber \\
+ \frac{1}{4}[7 - 10\ln 2
 - 6\ln^{2}2 - 6Li_{2}(1)](R-1) + O[(R-1)^{2}]\}.
\end{eqnarray}
The real parts of $R_{A}^{i}(1,1)$ contribute to $\sigma_{A}$ in a product
together with $|\chi|^{2} \sim |R-1|^{-2}$. As a cosequence of their
proportionality to $(R-1)$ whose real part vanishes at the peak, the Z
exchange interference corrections become extremely small. For $\Delta
\neq $1, there are at the peak larger contributions due to ln$(R-1+\Delta)
\sim \ln\Delta$ in (~\ref{eq:r102},~\ref{eq:r103}).
 
A similar discussion applies to the $\gamma$Z interference corrections.
Due to (~\ref{eq:r41}), half
the $R_{A}^{i}(1,1)$ are combined there with the resonating function $\chi$
wich becomes imaginary at $R=1$. So, the imaginary parts of $R_{A}^{i}(1,1)$
are relevant. Again, the QED corrections are suppressed compared to the
Z exchange Born cross section behaving like $|1-R|^{-2}$.
 
So,  all  the interference contributions are small  since  the
photonic corrections $R_{A}^{i}(0,0)$ are also
non-resonanting at the peak.  In case of cuts to the photon energy,
$\Delta <$ 1, the \oalf
interference bremsstrahlung has to be taken into account properly.
For  very small $\Delta$, logarithms of the type  $\ln\Delta$  may
become  even dominating and then
one should try to apply some soft photon summation  procedure.
 
 
\vfill\eject
\vspace{1.cm}
       5.   FINAL STATE RADIATION
\vspace{1.cm}
\nopagebreak
 
 
\vspace{1.cm}
       5.1. THE RADIATOR FUNCTIONS FOR THE ANGULAR DISTRIBUTION
\vspace{1.cm}
\nopagebreak
 
 The  final state radiation contributions to the differential
cross section have a simple angular dependence  compared
to the expressions discussed in the  preceding sections.  The
soft  photon parts of the final state  radiators  $S_{A}^{f}
(c,\epsilon;m,n)$ may be
obtained from $S(\epsilon,\beta_{e})$ as defined in (~\ref{eq:s34})
by replacing $m_{e}$ by the final state mass $m_{f}$ :
\begin{equation}
 S_{A}^{f}(c,\epsilon;m,n) = D_{A}(c)\:S(\epsilon,\beta_{f}),
           \hspace{1cm}A = T,FB,
\label{eq:s51}
\end{equation}
where  $D_{A}$(c)  is  introduced  in  (~\ref{eq:d32},~\ref{eq:d33}).
The  hard  photon radiators are:
\begin{equation}
  H_{T}^{f}(v,c) = \frac{\alpha}{\pi}Q_{f}^{2}\{D_{T}(c)
                    [H_{f}(v)-3v]+4v\},
\label{eq:h52}
\end{equation}
\begin{equation}
  H_{FB}^{f}(v,c) = \frac{\alpha}{\pi}Q_{f}^{2}D_{FB}(c)h_{f}(v),
\end{equation}
where the radiators $H_{f}(v), h_{f}(v)$ have been derived earlier for
the integrated cross sections $\sigma_{T}$ and $\sigma_{FB}$
$^{\cite{pl:dyb}}$:
\begin{equation}
  H_{f}(v) = \frac{1+(1-v)^{2}}{v}[L_{f}-1+\ln (1-v)],
\end{equation}
\begin{equation}
  h_{f}(v) = \frac{2}{v}[(1-v)(L_{f}-1)+\ln (1-v) + \frac{1}{2}v^{2}L_{f}].
\label{eq:h55}
\end{equation}
The integration of the symmetric radiator function $(A = T)$  over
the  photon  energy gives a contribution to  the  differential
cross section:
\begin{equation}
 R_{T}^{f}(c;m,n) = \frac{1}{4}(5-3c^{2}).
\label{eq:r56}
\end{equation}
The corresponding anti-symmetric radiator integral vanishes:
\begin{equation}
 R_{FB}^{f}(c;m,n) = 0.
\label{eq:r57}
\end{equation}
As has been pointed out e.g.  in $^{\cite{yr:boh}}$
, the soft photon corrections (~\ref{eq:s51})
 to even and odd cross section parts are equal and
will, consequently, cancel for the forward backward asymmetry.
After integration over the full photon phase space, eqs.
(~\ref{eq:r56}-~\ref{eq:r57})
,  there is minor influence on \afb due to the correction  in
the denominator.  Nevertheless,  for small cut off values  $\Delta
\ll$ 1,  soft  photon exponentation is recommended in order to  get
reliable  numerical  results.
A naive addition of initial and final state radiation  corrections
as  has been performed in $^{\cite{pl:dyb}}$
changes this  dependence  and
leads to an overestimation of final state radiation due to
an inbalance of corrections. In order to maintain the smallness
of isolated final  state corrections after  combination  with
the  large contributions from the initial state,  one has  to
proceed more  carefully.  A treatment based on the intuitive picture
of  subsequent radiation of photons from initial and final states
is proposed in subsection 5.3.
An isolated soft photon exponentiation  for
final  state radiation is analogue to the procedure  described
in  $^{\cite{np:bbn}}$ and consists of the replacements
\begin{equation}
 \delta(v)S_{A}^{f}(c,\epsilon;m,n) \rightarrow \beta_{f}v^{\beta_{f}-1}
 D_{A}(c)\bar{S}(\beta_{f}),
\label{eq:d58}
\end{equation}
\begin{equation}
 H_{A}^{f}(v,c) \rightarrow \bar{H}_{A}^{f}(v,c) = H_{A}^{f}(v,c) -
 \frac{\beta_{f}}{v}D_{A}(c).
\label{eq:h59}
\end{equation}
 
Numerical results due to final state radiation are shown in fig.4.
 
\vspace{1.cm}
       5.2. INTEGRATED CROSS SECTION AND FORWARD BACKWARD ASYMMETRY
TO ORDER \oalf
\vspace{1.cm}
\nopagebreak
 
The  integrated  final  state  \oalf QED  corrections  are   simple:
\begin{equation}
 R_{A}^{f}(m,n) = \frac{\alpha}{\pi}Q_{f}^{2}\left[\frac{1}{2}(L_{f}-1)(4\ln
     \Delta+3-4\Delta+\Delta^{2})-2Li_{2}(\Delta)+2Li_{2}(1)+r_{A}^{f}
         (\Delta)\right],
\label{eq:r129}
\end{equation}
\begin{equation}
  r_{T}^{f}(\Delta) = \frac{1}{2}(1-\Delta)(3-\Delta)L_{1} +
              \frac{1}{4}(6\Delta-2-\Delta^{2}),
\end{equation}
\begin{equation}
  r_{FB}^{f}(\Delta) = \frac{1}{2}(\Delta^{2}-1).
\label{eq:r131}
\end{equation}
If the photon energy cut is removed, one gets:
\begin{equation}
  R_{T}^{f}(m,n) = \frac{\alpha }{\pi}Q_{f}^{2}
\frac{3}{4},
\end{equation}
\begin{equation}
  R_{FB}^{f}(m,n) = 0.
\end{equation}
Figs. 5 and 6 illustrate the above discussion of the corrections to the
angular distribution due to final state radiation.
A  comparison  of  (~\ref{eq:r57})  with
(~\ref{eq:h55},~\ref{eq:h59})  shows  that  the
contribution  of  final state radiation  to $\sigma_{FB}$ vanishes  if
neither  a cut on the photon energy is applied
$^{\cite{du:bbf}}$ nor  finite mass  effects are taken into account
$^{\cite{pr:jlz}}$. Then,  the only (and
minor) influence of final state radiation on \afb  is due to the
C-even correction in (~\ref{eq:a24}).
If a tight cut is applied ($\Delta \ll$ 1),
the C-even and C-odd final state corrections (~\ref{eq:r129}-
~\ref{eq:r131})  approach
each  other because from soft photons there is no influence on
the  angular  behaviour of the emitted fermions.  Due  to  the
logarithmic cut dependence,  one again should exponentiate  the
soft  photon  contributions in that case but we leave out  the
details here. The ansatz is defined in (~\ref{eq:d58}-~\ref{eq:h59}).
 
 
\vspace{1.cm}
       5.3. SOFT PHOTON EXPONENTATION
\vspace{1.cm}
\nopagebreak
 
As may be seen from (~\ref{eq:h52}-~\ref{eq:h55}), the final state radiation
contributions  to  the differential cross section behave as  a
function of the normalised photon energy $v$ quite similar to the integrated
quantities $\sigma_{A},
A = T,FB$.  Based on their simple dependence on the  scattering
angle, we now derive a common treatment of soft photon
exponentiation  for initial and final state corrections.
We start  from the definition (~\ref{eq:d21},~\ref{eq:r22}):
\begin{eqnarray}
 \frac{d\sigma^{(e+f)}}{dc} = \sum_{m,n=0,1} \sum_{A=T,FB} Re\left\{
     \int_{0}^{\Delta}\right.d v\: \sigma_{A}^{0}(s',s';m,n)
 \nonumber \\
\{\delta (v)[1+S(\epsilon ,\beta_{e}))D_{A}(c)]+\theta (v-\epsilon )
     H_{A}^{e}(v,c)\}
 \nonumber \\
 + \sigma_{A}^{0}(s,s;m,n) D_{A}(c)\left. \int_{0}^{\Delta}d u\:
     [\delta (u) S(\epsilon ,\beta_{f}) + \theta (u-\epsilon )
    \tilde{H}_{A}^{f}(u,c)]\right\},
\label{eq:s515}
\end{eqnarray}
\begin{equation}
  H_{A}^{f}(u,c) = D_{A}(c)\tilde{H}_{A}^{f}(u,c),
\end{equation}
where $\tilde{H}_{FB}^{f}$ is independent of the
scattering angle and $\tilde{H}_{T}^{f}$    has
an almost negligible dependence.
     Further, (~\ref{eq:s515}) can be rewritten as follows:
\begin{eqnarray}
 \frac{d\sigma^{(e+f)}}{dc} = \sum_{m,n=0,1} \sum_{A=T,FB} Re
     \int_{0}^{\Delta}d v\: \sigma_{A}^{0}(s',s';m,n)
 \nonumber \\
 \{\delta(v)[1+S(\epsilon,\beta_{e})+\tilde{r}_{A}^{f}(s',c,\Delta)]
    D_{A}(c)+\theta(v-\epsilon)H_{A}^{e}(v,c)\},
\end{eqnarray}
\begin{equation}
  \tilde{r}_{A}^{f}(s,c,\Delta) = \int_{0}^{\Delta}d u
  [\delta(u)S(\epsilon,\beta_{f}) + \theta(u-\epsilon)
  \tilde{H}_{A}^{f}(u,c)].
\end{equation}
The  dependence of $\tilde{r}_{A}^{f}$ on $c$ is very weak for
$A = T$ and  absent for $A = FB$.
     Now  the  following  ansatz  seems  to  include  a  quite
reasonable   approximation   of  higher  order  soft   photon corrections:
\begin{equation}
 \frac{d\bar{\sigma}^{(e+f)}}{dc} =
 \sum_{m,n=0,1} \sum_{A=T,FB} Re
     \int_{0}^{\Delta}d v \:\sigma_{A}^{0}(s',s';m,n)
    F_{A}^{e}(v,c)F_{A}^{f}(v,c,s'),
\end{equation}
\begin{equation}
  F_{A}^{e}(v,c) = \delta(v)D_{A}(c)[1+S(\epsilon,\beta_{e})]
  + \theta(v-\epsilon)H_{A}^{e}(v,c),
\end{equation}
\begin{equation}
  F_{A}^{f}(v,c,s') = 1 + \tilde{r}_{A}^{f}(s',c,\Delta).
\end{equation}
The  exponentiation of soft photon contributions leads  to  the
following replacements:
\begin{equation}
  F_{A}^{e}(v,c) \rightarrow \bar{F}_{A}^{e}(v,c)
  = D_{A}(c)\beta_{e}v^{\beta_{e}-1}[1+\bar{S}(\beta_{e})] + \bar{H}
_{A}^{e}(v,c),
\end{equation}
\begin{equation}
  F_{A}^{f}(v,c,s') \rightarrow \bar{F}_{A}^{f}(v,c,s')
  = \int_{0}^{\Delta '}d u\: \{\beta '_{f}u^{\beta_{f}'-1}[1+\bar{S}
    (\beta_{f}')] + \bar{\tilde{H}}_{A}^{f}(u,c,\beta_{f}')\},
\label{eq:f523}
\end{equation}
\begin{equation}
 \beta_{f}' = \frac{2\alpha}{\pi}Q_{f}^{2}(\ln\frac{s'}{m_{f}^{2}}
   -1).
\end{equation}
Here again the residual hard radiator parts $\bar{H}$ are
\begin{equation}
 \bar{\tilde{H}}^{a}_{A}(v,c,\beta) = \tilde{H}^{a}_{A}
 (v,c,\beta) -\frac{\beta}{v}D_{A}(c).
\end{equation}
The  integrations  over  the variable u can  be  performed  in
(~\ref{eq:f523}):
\begin{equation}
 \bar{F}_{A}^{f}(v,c,s') = \Delta '^{\beta_{f}'}[1+\bar{S}(\beta_{f}')]
+ G_{A}(v,c,s'),
\end{equation}
\begin{equation}
 G_{A}(v,c,s') = \frac{1}{4}\beta_{f}'\Delta '(\Delta '-4) -
\frac{\alpha}{\pi}Q_{f}^{2}[2Li_{2}(\Delta ') + g_{A}(\Delta ',c)],
\end{equation}
\begin{equation}
g_{T}(\Delta ',c) = \frac{1}{2}(1-\Delta ')(3-\Delta ')\ln(1-\Delta ')
  - \frac{1}{4}\Delta '(\Delta '-6) - \left(3-\frac
 {4}{1+c^{2}}\right)\frac{\Delta '^{2}}{2}.
\end{equation}
\begin{equation}
  g_{FB}(\Delta ',c) = \frac{1}{2}\Delta '^{2},
\end{equation}
In the above formulae, we introduced the variable $\Delta '$ which is
connected with the reduced $s'$ in the same way as $\Delta$ with $s$:
\begin{equation}
  \Delta ' \equiv 1 - \frac{s_{min}'}{s'},
\end{equation}
\begin{equation}
  \Delta ' = \frac{\Delta-v}{1-v}.
\end{equation}
if  one  wants  to obtain from the above expressions  for  the
angular  distribution those for \st and \afb ,
  it is helpful  to
start  with  the observation that the $\bar{F}_{A}^{f}$
are independent  of  the
scattering  angle  with  exclusion of $g_{T}(\Delta ',c)$.
 Since  that
dependence is minor, i.e. less than $\frac{1}{2}\frac{\alpha}{\pi}Q_{f}^
{2} = 0.1 \%  Q_{f}^{2}$ ,
one can neglect it.  Then the final state radiation factor  is
completely   independent  of  the  scattering  angle  and  the
integration over c has to be done for F$_{A}^{e}$(v,c) only:
\begin{equation}
  \bar{\sigma}_{A}^{(e+f)} = \sum_{m,n=0,1}
  Re \int_{0}^{\Delta}d v \sigma_{A}^{
0}(s',s';m,n)\bar{F}_{A}^{e}(v)\bar{F}_{A}^{f}(v,s'),
\label{eq:s531}
\end{equation}
\begin{equation}
  \bar{F}_{A}^{e}(v) = \frac{3}{4}\int_{0}^{1}d c F_{A}^{e}(v,c).
\end{equation}
This integration yields the normal convolution  representation
for  the  total cross section $\bar{\sigma}_{A}$, but now modified
by the additional factor $\bar{F}_{A}^{f}$.
  So,  by a  direct  Feynman
diagram  calculation  we have shown that (~\ref{eq:s531})
 is  an
approximation  of sufficient accuracy though some well-defined
terms of the order \oalf have been neglected. A similar formula
in the case of \st
has been obtained by other methods in $^{\cite{pl:nit}}$.
The dependence of \st and \afb on the cut-off parameter $\Delta$
is shown in figs. 5 and 6 for different handlings of initial
and final state corrections. For smaller $\Delta$ the cross section becomes
smaller since the positive hard photon part is restricted then.
In the tail region (figs. 5c, 6c), one observes a steeply falling cross
section at values of $\Delta$ which cut away the radiative tail.
This happens if the effective energy $s'$ cannot reach the resonance
energy, i.e. $\Delta \leq$ 1-M$_{Z}^{2}$/s.
Even for
infinitesimal $\Delta$ the net negative bremsstrahlung corrections remain
finite due to soft photon exponentiation as discussed in the text. In
the figures, we applied the simplified but sufficiently accurate procedure
explained in chapter 4.2. Since the corresponding soft photon corrections
in units of the Born cross sections \st and $\sigma_{FB}$ are equal,
the net final state correction to \afb vanishes for vanishing $\Delta$.
For sufficiently large cut-off values, the final state contributions
also remain nearly negligible. In the intermediate region, they must  be
taken into account at least for precision measurements.

\vspace{1.cm}
       6. DISCUSSION
\vspace{1.cm}
\nopagebreak
 
In the foregoing chapters, basic formulae for and characteristic features
of the QED corrections to three different observables in
fermion pair production from \ee annihilation have been presented separately.
Numerical results have been produced with the codes MUCUT and MUCUTCOS
 of the package ZBIZON$^{\cite{zb:bar}}$.
 
In fig. 7, the net QED corrections to the differential cross section
are shown in combination with the weak loop corrections.
The latter have been determined with the code DIZET which is also part
of ZBIZON.
 For $M_{Z}$=91.1,
$m_{t}$=100.,$M_{H}$=100. (all masses in GeV), and $\alpha_{s}$=0.12 we get
in accordance with $^{\cite{np:abr}}$$^{\cite{yr:fab}}$: sin$^{2}\Theta_{W}$=
0.2314,$\Gamma_{Z}$=2.477 GeV. The corrected muon decay constant, running
QED coupling, and weak neutral vector and axial vector couplings to be
used in (~\ref{eq:c14},~\ref{eq:c17},~\ref{eq:tr1},~\ref{eq:tr2}) are
$^{\cite{cp:bar}}$:
\begin{equation}
 \alpha F_{A}(M_{Z}^{2}) = \alpha(1.063 - i 0.018),
\end{equation}
\begin{equation}
 G_{\mu}\rho_{Z}(M_{Z}^{2}) = G_{\mu}(1.000 - i 0.005),
\end{equation}
\begin{equation}
 a_{e} = -\frac{1}{2},
\end{equation}
\begin{equation}
 v_{e}(M_{Z}^{2}) = -\frac{1}{2}[1 - 4\sin^{2}\Theta_{W}(1.011 +i 0.013)],
\end{equation}
\begin{equation}
 v_{ee}(M_{Z}^{2}) =\frac{1}{4}[1 - 8\sin^{2}\Theta_{W}(1.011 + i 0.013)
 + 16\sin^{4}\Theta_{W}(1.023 + i 0.027)].
\end{equation}
If, for the sake of comparison, the initial-final interference terms
are excluded, the fully corrected total cross
section \st  as obtained here agrees within $\pm 0.2\%$ with that obtained
from the code ZSHAPE $^{\cite{np:bbn}}$$^{\cite{yr:fab}}$ which is
exact to order O($\alpha^{2}$). For \afb and d$\sigma$/dcos$\Theta$
, we do not know of a program which would allow for a similar
comparison.
 
While the weak loop corrections remain small $^{\cite{yr:fab}}$
$^{\cite{yr:boh}}$, the QED corrections amount typically to several
percents or more, essentially due to initial state radiation. Even
in the tail region, the angular dependence of the corrected differential
cross section follows closely the Born cross section behaviour with
exclusion of scattering angles near
                    $|\cos\Theta|=1$. This is reflected also
in figs. 8 and 9 where \st and \afb are shown as functions of an acceptance
cut, $|\Theta| < \Theta_{max}$. In general, the asymmetry is not a monotonic,
rising with $\Theta_{max}$ function though there are regions with an
almost linear behaviour. For loose cut values, the tail effect is
pronounced at $s > M_{Z}^{2}$. The total asymmetry is small at resonance
with corrections nearly of the order of the Born contribution. The
relatively large asymmetry value for a tight cut, $E_{\gamma} < $1GeV,
may be understood from figs. 2b where a systematic distortion of the
differential cross section is present in that case.
 
In some recent determinations of \st $^{\cite{pl:lep}}$ and
\afb $^{\cite{l3:ade}}$ at LEP, the QED corrections $^{\cite{pr:cbc}}$
$^{\cite{np:bbn}}$$^{\cite{zb:bar}}$ have been applied to data corrected
to correspond to a detector with a full angular acceptance (4$\pi$ geometry).
From figs. 8,9 one can see that this may be justified although the present
formulae allow a refined analytic approach to the data. This seems to be
recommended at least for the interpretation of precision measurements.
 
In view of the relative smallness of interference and final state corrections,
it is interesting to know where they can be neglected completely. Of
course, this depends on the energy region and  photon energy
cut-off chosen. In figs. 10,11 corresponding regions of relevance are shown.
 
To summarise, the analytic approach to QED corrections has been proven
quite powerful both for consistency checks of Monte Carlo programs
$^{\cite{yr:kle}}$ and for the interpretation of experimental data. This
article contains a first, systematic presentation of our analytic formulae
for the complete set of QED corrections to the differential cross
section d$\sigma$/dcos$\Theta$, total cross section \st and integrated
forward backward asymmetry \afb.
 
\vspace{1.cm}
      ACKNOWLEDGMENTS
\vspace{1.cm}
 
 We would like to thank  F.A. Berends, M. Greco, W.Hollik, R.Kleiss
and Yu. Sedykh  for
enligthening discussions and fruitful cooperation.

\vfill\eject
\vspace{1.cm}
            FIGURE CAPTIONS
\vspace{1.cm}
\nopagebreak
 
Figure caption
Fig. 1. The C-even and C-odd contributions to d$\sigma$/d$\cos\Theta$ due to
Z exchange (solid line), $\gamma$ exchange (dashed line),
$\gamma$Z interference
(dotted line) in nbarn as functions of the scattering angle
at $\sqrt{s}$ =30 GeV (a)
91.1GeV (b), 200 GeV (c).
Parameters: $M_{Z}$ = 91.1 GeV, $\Gamma_{Z}$ =2.5GeV,$\sin^{2}\Theta_{W}$=.23,
$\Delta$=1.
 
Fig. 2. The differential cross section with Born and initial state radiation
contributions in nbarn as functions of the scattering angle and of the photon
energy cut-off $\Delta$.  Other parameters: see fig.1.
 
Fig.3. QED contribution to d$\sigma$/d$\cos\Theta$ due to initial-final
interference as function of the scattering angle and $\Delta$.
Parameters: see figs.1,2.
 
Fig. 4. Born plus final state correction contributions to
d$\sigma$/d$\cos\Theta$ in nbarn as function of the scattering angle and $\Delta$.
Parameters: see figs.1,2.
 
Fig. 5. The total cross section \st as function of the photon energy
cut-off $\Delta$.
Parameters: see figs.1,2.

Fig. 6. The integrated forward-backward asymmetry \afb as function of
the photon energy cut-off $\Delta$.
Parameters: see figs.1,2.
 
Fig. 7. The differential cross section with complete electroweak
corrections as function of the scattering angle and $\Delta$.
Parameters $M_{Z}$=91.1, $m_{t}$=100.,$M_{H}$=100. (all masses in GeV),
$\alpha_{s}$=0.12.
 
Fig. 8.The total cross section \st as function of an acceptance cut on the
scattering angle, $|\cos\Theta|\le \cos\Theta_{max}$.
Parameters see fig. 7.
 
Fig. 9. The integrated forward-backward asymmetry \afb as function
of an acceptance cut on the scattering angle.
Parameters see fig. 7.

Fig. 10. Cross section contributions due to initial-final state interference
(a) and final state radiation (b) in percent as functions of $\sqrt{s}$ and
$\Delta$.
Parameters: see figs.1,2.

Fig. 11. Asymmetry contributions due to initial-final state interference
(a) and final state radiation (b).
Parameters: see fig. 10.
 
\vfill\eject

\end{document}